\documentstyle[preprint,prc,aps]{revtex}
\newcommand{\hic}{heavy-ion collision }
\newcommand{\hics}{heavy-ion collisions }
\newcommand{\nnc}{nucleus-nucleus collision }

\newcommand{\apr}{antiproton }
\newcommand{\aps}{antiprotons }

\newcommand{\pncs}{proton-nucleus collisions }
\newcommand{\se}{selfenergy }
\newcommand{\ses}{selfenergies }

\newcommand{\sep}{Schrdinger-equivalent potential }

\newcommand{\barp}{$\bar{p}$}
\newcommand{\anni}{annihilation }
\newcommand{\prob}{probability }
\newcommand{\pnr}{proton-nucleus reaction }
\newcommand{\pnrs}{proton-nucleus reactions }
\newcommand{\crs}{cross section }
\newcommand{\crss}{cross sections }
\newcommand{\abs}{absorption }

\newcommand{\be}{\begin{equation}}
\newcommand{\ee}{\end{equation}}
\newcommand{\bea}{\begin{eqnarray}}
\newcommand{\eea}{\end{eqnarray}}

\begin{document}
\title{Analysis of Subthreshold
Antiproton Production in p-Nucleus and Nucleus-Nucleus Collisions
in the RBUU Approach
    \protect \footnote[2]{Supported by BMFT, GSI Darmstadt and KFA J\"ulich.} }
\author{Stefan Teis, Wolfgang Cassing, Tomoyuki Maruyama \\
	    and Ulrich Mosel \\
	Institut f\"{u}r Theoretische Physik, Universit\"{a}t Giessen \\
	D-35392 Giessen, Germany}
\maketitle
\begin{abstract}
We calculate the subthreshold production of antiprotons in the
Lorentz-covariant RBUU approach employing
a weighted testparticle method to treat the \apr propagation
and absorption nonperturbatively. We find that the \apr differential
cross sections are highly sensitive to the baryon and antiproton
selfenergies in the dense baryonic environment. Adopting the baryon
scalar and vector selfenergies from the empirical optical potential
for proton-nucleus elastic scattering and from
Dirac-Brueckner calculations at higher density $\rho > \rho_0$ we
examine the differential \apr spectra as a function of the
antiproton selfenergy.
A detailed comparison with the available experimental data
for p-nucleus and nucleus-nucleus reactions  shows
that the antiproton feels a moderately attractive mean-field at normal
nuclear matter density $\rho_0$ which is in line with a dispersive
potential extracted from the free annihilation cross section.

\end{abstract}

\newpage
\section{Introduction}
The production of particles at energies below the free nucleon-nucleon
threshold ('subthreshold production') constitutes one of
the most promising sources of information about
the properties of nuclear matter at high densities since the particles are
produced predominantly during the compressed stage at high density
\cite{Bertsch88,Cassing90,mosel,Aich}.
Antiproton production at energies of a few GeV/u is the most
extreme subthreshold production process and has been
observed in proton-nucleus collisions already more than 20 years ago
\cite{chamberlain,elioff,dorfan}.
Experiments at the JINR \cite{JINR} and at the BEVALAC
\cite{BEVALAC1,BEVALAC2} have provided, furthermore, first measurements of
subthreshold antiproton production in nucleus-nucleus collisions. Since then
the problem was taken up again at KEK \cite{KEK} and GSI \cite{GSI}
with new detector setups. Various
descriptions for these data have been proposed.
Based on thermal models it has been suggested that the antiproton yield
contains large contributions from $\Delta N \rightarrow \bar{p} +X$,
$\Delta \Delta \rightarrow \bar{p} + X$  and $\rho \rho \rightarrow \bar{p}N$
production mechanisms \cite{Koch89,ko1,ko2}.
Other models have attempted to explain these data in terms of multiparticle
interactions \cite{Danielewicz90}.

In a first chance nucleon-nucleon
collision model (assuming high momentum tails consistent with data on backward
proton scattering) Shor et al. \cite{Shor90} succeeded in reproducing the
 antiproton yield  for the proton-nucleus case; however, these authors
 underestimated the yield by more than 3 orders of magnitude for
nucleus-nucleus collisions. This problem was partly resolved by Batko et
al. \cite{Batko91} who performed the first nonequilibrium $\bar{p}$-production
study on the basis of the VUU transport equation.
Essential for this success was that in
A + A reactions the dominant production channel proceeds via an
intermediate nucleon resonance which allows to store a sizeable amount of
energy that can be used in a subsequent collision for the production of a
$ p \bar{p}$ pair. Later on, these findings were also
confirmed by Huang et al. \cite{Faessl}.

These results have led to the suggestion that the quasi-particle
properties of the nucleons might be important for the $\bar{p}$ production
process which become more significant with increasing nuclear density.
Schaffner et al. \cite{Schaffner91} found in a static thermal relativistic
model, assuming
kinetic and chemical equilibrium, that the $\bar{p}$-abundancy might be
dramatically enhanced when assuming the antiproton selfenergy to be
given by charge conjugation of the nucleon selfenergy. This leads to
strong attractive vector selfenergies for the antiprotons. However, the above
concept lacks unitarity between real and imaginary part of the \barp-\se and
thus remains questionable. Besides this, even in the $\sigma-\omega$-model the
Fock terms lead to a suppression of the attractive \barp-field
\cite{Soutome90}, such that the production threshold is shifted up in
energy again as compared to the simple model involving charge conjugation.
Furthermore, the assumption of thermal and especially chemical equilibrium
most likely is not fullfilled e.g. in $Si+Si$ collisons around 2 GeV/u
\cite{Lang90}.

First preliminary results of a fully relativistic transport calculation
for antiproton
production including $\bar{p}$ annihilation as well as the change of the
quasi-particle properties in the medium have been reported in \cite{Cass92}.
There it was found that according to the reduced nucleon mass in the
medium the threshold for $\bar{p}$-production is shifted to lower energy
and the antiproton cross section prior to annihilation becomes enhanced for
$Si+Si$ at 2.1 GeV/u by approximately a factor 70 as compared to a
relativistic cascade calculation where no in-medium effects are incorporated.
However, all these transport calculations \cite{Batko91,Faessl,Cass92}
suffered from an approximate geometrical
treatment of the very strong annihilation channel
and a neglect of the momentum dependence of the baryon selfenergies.

In our present work we therefore analyze the production of antiprotons in
the framework of the relativistic transport theory (RBUU) where
antiprotons are propagated explicitly in the respective time dependent
potentials and their annihilation is calculated nonperturbatively
by means of individual rate equations.

First results have been published in \cite{Teis93}. Results of a
very similar investigation by Li et al. \cite{Ko93} have recently become
available and we compare frequently with the outcome of this study.

In this respect we first discuss two simple models for the $\bar{p}$ selfenergy
in section 2 and present the general transport equations for nucleons and
antinucleons in section 3. The numerical implementation of $\bar{p}$ production
is presented in section 4 as well as a detailed analysis of the production
process with respect to the selfenergies employed, the systematics with
respect to projectile and target masses and the individual baryonic
production channels. In section 5 we present the treatment of $\bar{p}$
propagation and annihilation and discuss the geometrical aspects of $\bar{p}$
absorption. The explicit comparison of our calculations with the available
experimental data for $p + A$ and $A + A$ reactions is performed in section 6,
while a summary on the $\bar{p}$ selfenergies concludes the paper in
section 7.

\section{Models for the antiproton selfenergy}
\label{antise}
The dynamics and the properties of particles in a many-body system
strongly depend on their mutual interactions with the surrounding
particles and are reflected in their selfenergies.
While the real part of the selfenergy describes the change of the
particle momenta in the medium, all inelastic
reaction channels as well as elastic scattering processes are accounted for
by the imaginary part of
the selfenergy $\Sigma(p_\mu,\rho_s,j_\mu)$. If the selfenergy $\Sigma$
is an analytic function of the
particle energy $\epsilon = p_0$ its imaginary and real part are related up to
a constant by the dispersion relations:
\bea
Re \left[ \Sigma(\epsilon) \right]	& = & \frac{1}{\pi} P \int_{0}^{+
\infty}
	\frac{Im \left[\Sigma\left( \epsilon' \right) \right]}{\epsilon' -
	\epsilon} d\epsilon'
	    \label{dispses1}\\
Im \left[ \Sigma(\epsilon) \right]	& = & -\frac{1}{\pi} P \int_{0}^{+
\infty}
	\frac{ Re \left[\Sigma\left( \epsilon' \right) \right]}{\epsilon' -
	\epsilon} d\epsilon'.
\label{dispses}
\eea
Since we assume particles and antiparticles to be different particle
species the integration in the Principal-value integrals (\ref{dispses1}) and
(\ref{dispses}) extends only over positive energies.

In the relevant energy regime (1 - 5 GeV/u) of \hics and \pncs
the \ses of baryons and \aps are known to be quite different. The
\barp-\se is dominated by the \apr annihilation with baryons
($\bar{p}+B \rightarrow X$) as shown by low energy antiproton-nucleon
 scattering experiments at LEAR \cite{Walcher89,Amsler91}.
Since the \apr production probability at subthreshold energies is extremly low
the number of \aps produced in \hics and \pncs is negligible
compared to the number of baryons involved in the reaction.
Thus contributions from elastic \barp-\barp-scattering to
the \apr \se can be neglected as well as  contributions
of \barp-\anni channels to the baryon \ses.

\subsection{A dispersive model}
%simple model
In the following we present a simple model for the real part of the
\barp-\se using the dispersion relation
(\ref{dispses1}). In the low density limit
the imaginary part of the \se (neglecting elastic scattering) is given
 by the integral :
\be
 2 \: Im \, \Sigma_{\bar{p}}(x,\, \Pi_{\bar{p}})
 = - \frac{4}{(2 \pi)^{3}} \int
d^{3} \Pi_{B} \, \frac{m^{\star}_{B}}{\Pi_{B}^{0}} W(\Pi_{\bar{p}}^{\mu},
\Pi_{B}^{\mu}) f(x,\Pi_{B})
\label{cass4}
\ee
where  $W(\Pi^{\mu}_{\bar{p}},
\Pi_{B}^{\mu})$ is the probability of an \apr with momentum
$\Pi^{\mu}_{\bar{p}}$ to annihilate in a collision with a
baryon of effective mass $m^{\star}$ and momentum $\Pi_{B}^{\mu}$.
$f(x,\Pi_{B})$ is the
baryon phase-space distribution function at space-time $x$ whereas the
factor 4 in formula (\ref{cass4}) arises from the summation over
the spin- and isospin-degrees of freedom. We simplify this expression
by considering an \apr moving in infinite nuclear matter of density
$\rho_{0}$ ($= 0.17 1/fm^{3}$) at rest, i.e.
$\vec{\Pi}_{B} = 0$. Both the \aps and the baryons now are supposed to
have effective masses equal to the restmass of the nucleon. This
leads to the following analytic expression for the \barp-\anni
\prob
\be
W(\Pi_{\bar{p}}^{\mu}, \Pi_{B}^{\mu}) = \frac{\sqrt{\Pi_{\bar{p}}^{0 \: 2} -
m_{N}^{2}}}{\Pi^{0}_{\bar{p}}} \cdot \sigma_{abs}(\Pi^{0}_{\bar{p}})
\label{cass0}
\ee
where $\sigma_{abs}$ is the total \barp-\anni cross section. Due to the above
assumptions the integral over the baryon phase-space distribution
function and the summation over the spin- and isospin-degrees of
freedom in eq. (\ref{cass4}) can be replaced by a multiplication of the
 \anni \prob with $\rho_{0}$. The expression for the imaginary part of
 the \apr \se as a function of the \apr energy then reads
\be
Im \,  \Sigma_{\bar{p}}(\Pi_{\bar{p}}^{0}) = -\frac{1}{2}
\frac{\sqrt{\Pi_{\bar{p}}^{0 \: 2} -
m_{N}^{2}}}{\Pi^{0}_{\bar{p}}} \cdot \sigma_{abs}(\Pi^{0}_{\bar{p}})\,
\rho_{0}
\label{cass5}
\ee
and the real part of $\Sigma$ at density $\rho_0$ can be calculated
from eq. (\ref{cass4}) by employing the dispersion relation (\ref{dispses1}).
We expect this expression to give a reasonable description of $Im
\Sigma_{\bar{p}}$ at least for densities up to $\rho_{0}$ where
possible medium corrections to $\sigma_{abs}$ are still small.

In order to estimate $Re \Sigma(\epsilon)$ at $\rho_0$
we adopt the parametrization from \cite{Koch89}
for $\sigma_{abs}$, i.e.
\be
\sigma_{abs}  =  \sigma_{0} \frac{s_{0}}{s} \left(
		    \frac{A^{2} \, s_{0}}{(s-s_{0})^{2} + A^{2}\, s_{0}}
		    + B \right)
		    \label{para}
\ee
with the constants $\sigma_{0}      =  120 \: mb$, A = 50 MeV, B = 0.6 and
$s_{0}=4m_{N}^{2}$. Fig. 1 displays this parametrization in comparison to
the experimental data for the free cross section taken from
\cite{annihilationdata} as a function of $s-s_{0}$ where s is the invariant
energy squared. The result of this calculation for $Re \Sigma(\epsilon)$
is displayed in fig. 2 as a function of the \barp-kinetic energy. In order
to investigate the sensitivity of this result to the parametrization of the
absorption cross section we used in addition
different parameters B (cf.  fig. 2) to vary the overall size of
$\sigma_{abs}$. For an \apr at rest we find values for the real part
of the \se of $-175$ MeV up to $-100$ MeV. While we observe a steep
rise for small kinetic energies this flattens out for higher
energies and reaches an asymptotic value of $0$ -
$-10$ MeV. Furthermore, the variation in the cross
section mainly affects the low energy regime.
The \barp-\ses obtained in the framework of this simple model
are well in line with the potential analysis for \barp  + A reactions
by Janouin et al \cite{Janouin86}.\\

\subsection{Mean-field models}

We now turn to the treatment of the \apr \ses in mean-field models where
the imaginary part of the \ses is not included at all thus
inherently violating the dispersion
relations (\ref{dispses1}, \ref{dispses}).

As an example we consider the mean-field approximation
of the familiar $\sigma-\omega$-model \cite{Serot} where
only momentum independent  vector $U^{\mu}_{v}(x)$ and scalar parts $U_{s}(x)$
 of the \ses are taken into account.
The equation of motion for a fermion spinor then reads
\be
\left\{ \gamma^{\mu} \left( i \partial_{\mu} - U_{\mu}^{v} (x) \right) -
\left( m + U^{s} \left( x \right) \right) \right\} \Psi \left( x
\right) = 0   \label{swdi1}
\ee
while the equation of motion for antiparticles is obtained by applying
the charge conjugation operator to this equation.
As a consequence the scalar part of
the \se is the same for particles and antiparticles while
the vector part of the antiparticle potential changes the sign
\bea
U^{v}_{C \, \mu} \left( x \right) & = & - U^{v}_{\mu} \left( x \right)
\label{uvc} \\
U^{s}_{C} \left( x \right) & = &  U^{s} \left( x \right).
\label{usc}
\eea
The combination of both parts of the antiparticle \se to the
\sep in the non-relativistic limit leads to a strongly attractive
potential for the antiprotons  when using the
original coupling constants from \cite{Serot}. The value of the
Schr\"odinger-equivalent potential at $\rho = \rho_{0}$
is approximately $-700$ MeV for
$\epsilon = 0$ and becomes even more attractive with
increasing kinetic energy of the \apr ($-1300$ MeV for $\epsilon = 2000$ MeV).

A comparison of the results for the \sep obtained in the framework of
the $\sigma - \omega$ model and the real part of the \barp-\se resulting from
the dispersive approach shows strong differences not
only in the absolute values but also in their opposite behaviour as a function
of the kinetic energy.
The fact that the real part of the \apr \se cannot be described consistently
within different models urges us to treat it as a free parameter in our
transport model. We regard the determination of this parameter in comparison
with the experimental data as the central goal of our work.

\section{The RBUU transport approach}

In this section we give a brief description of the RBUU-model. First we
summarize the relevant equations determining the dynamics of baryons
and then describe the implementation of \aps in our approach.

\subsection{Baryon dynamics}

Since the covariant BUU approach has been extensively discussed in the
reviews \cite{Cassing90c,Blaettel} and in \cite{KLW1} we only recall
the basic equations and the corresponding quasi-particle properties
that are relevant for a proper understanding
of the results reported in this study.
The relativistic BUU (RBUU) equation with momentum-dependent mean-fields or
selfenergies is given by (for details see refs.
\cite{Cassing90c,Blaettel,KLW1})
\begin{equation}
\{ [ \Pi_\mu - \Pi_\nu ( {\partial}^{p}_{\mu} U^{\nu} ) +
m^{\star} ( {\partial}^{p}_{\mu} U_s ) ]	\partial^\mu_x	+
 [ - \Pi_\nu ( {\partial}^{x}_{\mu} U^{\nu} ) +
m^{\star} ( {\partial}^{x}_{\mu} U_s ) ] \partial^\mu_p \} f(x,p) = I_{coll} ,
\label{090}
\end{equation}
where $f(x,p)$ is the Lorentz covariant phase-space distribution function,
$I_{coll}$ is a collision term (cf. eq. (\ref{icoll})), and $U_s$ and
$U_\mu$ are the scalar- and the vector- selfenergies.
The effective mass $M^*$ and the kinetic momentum $\Pi_\mu$ are defined in
terms of the fields by
\bea
\Pi_{\mu}(x,p) & = & p_{\mu} - U_{\mu}(x,p) \label{effp}\\
m^{\star}(x,p) & = & m + U_{s}(x,p) \label{effm} \\
\eea
while the quasi-particle mass-shell constraint is obtained as
\be
V(x,p)\,f(x,p) = 0
\label{e400}
\ee
with the pseudo potential
\be
V(x,p) \equiv \frac{1}{2} \{ \Pi^2 (x,p) - m^{*2} (x,p) \}.
\label{pseudo}
\ee
The above equation implies that the phase-space distribution function
$f(x,p)$ is nonvanishing only on the hypersurface in phase-space defined
by $V(x,p) = 0$. \\

In order to implement proper \ses for the nucleons in line with elastic
proton-nucleus scattering we follow ref.
\cite{KLW1} and separate the mean-fields into a local part and an
explicit momentum-dependent part, i.e.
\bea
U_{s}(x,p) & = & U_{s}^{H}(x) + U_{s}^{MD}(x,p) \label{pot16} \\
U_{\mu}(x,p) & = & U_{\mu}^{H}(x) + U_{\mu}^{MD}(x,p), \label{kwdef}
\eea
where the local mean-fields are determined by the usual Hartree equation
\bea
U_s^{H}(x) & = & - g_s \sigma_H (x)
\nonumber \\
U_\mu^H (x) & = & g_v \omega^H_\mu (x)
\label{e660}
\eea
with
\bea
m_{s}^2  \sigma_H (x) + B_{s}  \sigma_H^2 (x) + C_{s} \sigma_H^3 (x) & = &
g_{s}\rho_{s}(x)\nonumber \\
m_{v}^{2} \omega^H_\mu (x)  & = & g_{v} j_{\mu} (x) \label{sigom} .
\eea
In the above equations the scalar density $\rho_{s}(x)$ and the baryon
current $j_{\mu}(x) $ are given in terms of momentum integrals over
the phase-space distribution function (cf. \cite{KLW1,KLW2,Tomo}). The
potentials (\ref{kwdef}) and (\ref{pot16}) correct the unphysical
strong repulsion in the $p + A$ and $A+A$ potentials at high erngergies,
obtained in the original Walecka model. Refs. \cite{KLW1,KLW2} give
parametrizations which describe both the experimental data and the
density-dependence obtained in Brueckner-calculations very well.

Now we turn to the discussion of the collision term $I_{coll}$ describing
the baryon-baryon collisions (cf. \cite{Bertsch88,Cassing90c,Mal}):
\bea
I_{coll} & = & \frac{4}{(2 \pi)^{5}} \int d^{3} \Pi_{1} \, d^{3} \Pi'_{1} \,
              d^{3} \Pi'  \frac{m^{\star} \,m^{\star}_{1}\, m'^{\star} \,
              m'^{\star}_{1}}{\Pi^{0}\, \Pi^{0}_{1}\,  \Pi'^{0}
              \,  \Pi'^{0}_{1}}
\nonumber \\
	 &   & W(\Pi^{\mu},\Pi^{\mu}_{1} \mid \Pi'^{\mu},\Pi'^{\mu}_{1})
	\delta^{4}(p^{\mu} + p^{\mu}_{1} - p'^{\mu} - p'^{\mu}_{1} )
\nonumber \\
         &   & ( f(x,\Pi') f(x,\Pi'_{1})(1 - f(x,\Pi))(1-f(x,\Pi_{1}))
\nonumber \\
         &   & -  f(x,\Pi) f(x,\Pi_{1})(1 - f(x,\Pi'))(1-f(x,\Pi'_{1}))).
\label{icoll}
\eea
This collision integral describes the change in the phase-space
distribution function $f(x,\Pi)$ due to the collision of two baryons
with effective masses $m^{\star}$, $m^{\star}_1$ and momenta $\Pi^{\mu}$,
$\Pi^{\mu}_1$, respectively. The two baryons in the final state of the
 reaction with masses $m'*$ and $m_1'*$ are labeled by their momenta
$\Pi'$ and $\Pi_1'$.
The $\delta-$function guarantees energy and momentum conservation in
the individual collision while
$ W(\Pi^{\mu},\Pi^{\mu}_{1} \mid \Pi'^{\mu},\Pi'^{\mu}_{1})$ denotes the
transition probability for this reaction which can be expressed in the CMS
of the colliding particles by the product of the relative velocity of both
colliding particles and the differential cross section for the reaction,
\be
W(\Pi^{\mu},\Pi^{\mu}_{1} \mid \Pi'^{\mu},\Pi'^{\mu}_{1}) =  \mid \!
\frac{
\vec{\Pi} }{\Pi^{0}}+ \frac{ \vec{\Pi}}{\Pi^{0}_{1}}
\!  \mid
 \frac{d\sigma}{d \Omega} \mid_{
\Pi^{\mu} + \Pi^{\mu}_{1} \rightarrow \Pi'^{\mu} + \Pi'^{\mu}_{1} }.
\ee
For the NN cross section we use the parametrization given by Cugnon
\cite{Cugnon81} and therefore neglect the possible in-medium corrections.
However, the corrections due to Pauli-blocking are
built in via the factors $1-f$.  We explicitly account for the following
baryonic channels:
\bea
N + N \rightarrow N + N & , & N + N  \rightarrow N + \Delta \nonumber \\
N + \Delta \rightarrow N + N & , & \Delta +  \Delta  \rightarrow
 \Delta + \Delta \nonumber
\eea
and propagate the $\Delta$'s  with the same \ses as the nucleons. For the
parametrizations of the cross sections of reactions including $\Delta$'s
see \cite{Bertsch88}.\\
\subsection{Antiprotons in RBUU}

The phase-space distribution function for the \aps $f_{\bar{p}}(x,p_{\bar{p}})$
is assumed to follow an equation of motion equivalent to eq. (\ref{090}),
however, with scalar and vector potentials of different strength, i.e.
\bea
U_{s}^{\bar{p}}(x) & = & - g_{s}^{\bar{p}} \sigma_H(x) \nonumber \\
U_{\mu}^{\bar{p}}(x) & = & g_{v}^{\bar{p}} \omega^H(x)_\mu \label{appot} .
\eea
Thus the effective mass and the effective momentum of  antiprotons are
given by
\bea
m^{\star}_{\bar{p}} & =  m + U_{s}^{\bar{p}}(x) & =
m - g_{s}^{\bar{p}}\sigma_H(x) \label{effapm}  \\
\Pi_{\mu}^{\bar{p}} & =  p_{\mu}^{\bar{p}} - U_{\mu}^{\bar{p}}(x) & =
p_{\mu}^{\bar{p}} - g_{v}^{\bar{p}} \omega^H(x)_\mu. \label{effquap}
\eea
According to the arguments given in section \ref{antise} the coupling
constants $g_{s}^{\bar{p}}$ and $g_{s}^{\bar{p}}$ are treated as free
parameters and will be determined in a comparison of our
calculations to experimental data (cf. section \ref{res}).
Since at subthreshold energies antiprotons can be produced only in a
very limited kinematical range we can work here with scalar and
vector parts of the \apr \ses that are not explicitly
momentum-dependent. We then obtain the following equation of motion for the
antiproton phase-space distribution function
\be
\left[ \frac{\Pi^{\mu}_{\bar{p}}}{\Pi^{0}}_{\bar{p}}
 \partial^{x}_{\mu} + \left( \Pi^{\nu}_{\bar{p}} F_{\mu\nu}^{\bar{p}}
+ m^{\star}_{\bar{p}} \left( \partial_{\mu}^{x} m^{\star}_{\bar{p}} \right)
\right) \partial^{\mu}_{\Pi_{\bar{p}}} \right] f_{\bar{p}}(x,\Pi_{\bar{p}}
) = I_{coll}^{\bar{p}}(x,\Pi_{\bar{p}})
\label{vlat}
\ee
with
\be
F^{\mu\nu}_{\bar{p}}  =  \partial^{\mu} U_{v}^{\bar{p} \,\nu}
- -  \partial^{\nu} U_{v}^{\bar{p} \,\mu}
		  \label{fmunu}
\ee
and the mass-shell constraint
\be
\left( \Pi^{2}_{\bar{p}} - m^{\star\, 2}_{\bar{p}} \right)
 \bar{f} \left(x,\Pi_{\bar{p}} \right)
 =   0. \label{atmass3}
\ee

The collision term $I_{coll}^{\bar{p}}$ (r.h.s. of eq. (\ref{vlat}))
includes i) a term $I_{elast}^{\bar{p}}$, describing the elastic
\apr scattering, ii) a term $I_{prod}^{\bar{p}}$, describing the
\apr production, and iii) a term $I_{abs}^{\bar{p}}$
responsible for in-medium \barp-absorption.
$I_{elast}^{\bar{p}}$ describes elastic baryon-antiproton scattering
as well as elastic antiproton-antiproton scattering. While this part of the
collision integral $I_{coll}^{\bar{p}}$ can be formulated similarly to
the collision term describing bayon-baryon scattering (\ref{icoll})
the other terms represent extensions to this integral.

\subsubsection{Collision term for \barp-production}
The basis for the description of the \apr production is the reaction
\be
 B+B\rightarrow \bar{p}+p+N+N \equiv 1 + 2 \rightarrow
\bar{p} + 3+ 4+ 5, \label{prore1}
\ee
for which the corresponding covariant collision integral reads
\bea
\lefteqn{I^{\bar{p}}_{prod}(x,\Pi_{\bar{p}}) =} \nonumber \\
 &&  \frac{4}{(2 \pi)^{11}}
                     \int d^{3} \Pi_{1} \, d^{3} \Pi_{2} \,
                 d^{3} \Pi_{3} \, d^{3} \Pi_{4} \, d^{3} \Pi_{5}
  \frac{m^{\star}_{1} \, m^{\star}_{2} \, m^{\star}_{3}\,  m^{\star}_{4}\,
m^{\star}_{5}\,  m^{\star}_{\bar{p}}}{\Pi^{0}_{1}\, \Pi^{0}_{2}\,
\Pi^{0}_{3}\,  \Pi^{0}_{4}  \,
                 \Pi^{0}_{5} \,  \Pi^{0}_{\bar{p}} }\nonumber \\
         &&  W(\Pi^{\mu}_{1},\Pi^{\mu}_{2} \mid \Pi^{\mu}_{3},
                \Pi^{\mu}_{4},\Pi^{\mu}_{5},\Pi^{\mu}_{\bar{p}} )
	\delta^{4}(p^{\mu}_{1} +p^{\mu}_{2}- p^{\mu}_{3} -p^{\mu}_{4} -
  p^{\mu}_{5} -\Pi^{\mu}_{\bar{p}}  )
\nonumber \\
         && \left\{ f(x,\Pi_{1}) f(x,\Pi_{2})(1 - f(x,\Pi_{3}))
         (1-f(x,\Pi_{4})) (1 - f(x,\Pi_{5})) \right\} .
\label{collpr}
\eea
We have omitted the Pauli-blocking factor for the \apr in the final state
because the number of \aps created during a heavy-ion collision
in the subthreshold energy regime is negligible. For the same reason we
 neglect the effects of reaction (\ref{prore1}) on the phase-space
distribution function of the baryons.

\subsubsection{Collision term for \barp-absorption}

In the collision integral $I_{abs}^{\bar{p}}$ we do not treat all
possible annihilation reactions separately but sum up all channels in
the inclusive annihilation reaction
\bea
B + \bar{p}  \rightarrow  X,  \label{absrea1}
\eea
where X denotes all possible final states (essentially pions)
of the baryon-antiproton
annihilation. The corresponding energy and momentum conservation reads
\be
p_{B}^{\mu} + p_{\bar{p}}^{\mu} = p^{\mu}_X, \label{enmomabs}
\ee
where $ p_{B}^{\mu}$ and $ p_{\bar{p}}^{\mu}$ denote the 4-momenta of the
baryon and the antiproton, respectively, and $p^{\mu}_{X}$ stands for the
sum of the 4-momenta of all particles in the final state of the
annihilation reaction. Reaction (\ref{absrea1}) then leads to the
collision term
\bea
I_{abs}^{\bar{p}}(x,\Pi_{\bar{p}}) & = & - \frac{4}{(2 \pi)^{3}}
                     \int d^{3} \Pi_{1}  d^{4} \Pi_{X}
                   \frac{m^{\star}_{1}\,  m^{\star}_{\bar{p}}}{\Pi^{0}_{1}
        \,     \Pi^{0}_{\bar{p}} }\nonumber \\
	 &   &	W(\Pi^{\mu}_{1},\Pi^{\mu}_{\bar{p}} \mid \Pi^{\mu}_{X})
	\delta^{4}(p^{\mu}_{1} +p^{\mu}_{\bar{p}} - p^{\mu}_{X})
	 	f(x,\Pi_{1}) f(x,\Pi_{\bar{p}})
         \label{collabs3}
\eea
with $W(\Pi^{\mu}_{1},\Pi^{\mu}_{\bar{p}} \mid \Pi^{\mu}_{X})$ denoting the
transition probability for the reaction (\ref{absrea1}). Integrating
(\ref{collabs3}) over $d^4\Pi_X$ implies in addition to the integration
over all final momentum states of a particular reaction a summation
over all possible annihilation channels
\begin{equation}
I_{abs}^{\bar{p}}(x,\Pi_{\bar{p}})  =  - \frac{4}{(2 \pi)^{3}}
   \int d^{3} \Pi_{1}  \frac{m^{\star}_{1}\,  m^{\star}_{\bar{p}}}{\Pi^{0}_{1}
	\,     \Pi^{0}_{\bar{p}} } W(\Pi^{\mu}_{1},\Pi^{\mu}_{\bar{p}} )
         f(x,\Pi_{1}) f(x,\Pi_{\bar{p}}),
         \label{collabs32}
\end{equation}
where $W(\Pi^{\mu}_{1},\Pi^{\mu}_{\bar{p}})$ denotes the probability of an
\apr with effective momentum $\Pi^{\mu}_{\bar{p}}$ annihilating with
a baryon with effective momentum $\Pi_{1}^{\mu}$.
\section{Antiproton production in RBUU}
\label{approd}
In this section we describe the numerical treatment of
\apr production in
our model and present a systematic analysis
of the \barp-production mechanism in \hics.

\subsection{Numerical implementation}

Since the production \prob for \aps is very small
the average time evolution of the \nnc is not affected and it is
justified to treat the \barp-production perturbatively
\cite{Cassing90,Cassing90c}. In this approach the \apr invariant
differential multiplicity is obtained by summing incoherently over all
baryon-baryon collisions and integrating over all residual degrees of
freedom. Assuming the \apr production to take place via reactions
of the type
\bea
 B+B\rightarrow \bar{p}+p+N+N \equiv 1 + 2 \rightarrow
\bar{p} + 3+ 4+ 5 \label{prore}
\eea
(B stands for either nucleon or $\Delta$) the
invariant multiplicity as a function of the impact parameter can be written
as
\bea
E_{\bar{p}}\frac{d^{3}P(b) }{d^{3}\Pi_{\bar{p}}} & = &      \sum_{BB coll}
 \int d^{3} \Pi'_{3} d^{3} \Pi'_{4} d^{3} \Pi'_{5}  \frac{1}
 {\sigma_{BB}(\sqrt{s})}
E'_{\bar{p}} \frac{d^{12}
\sigma_{BB \rightarrow \bar{p}+X}
\left( \sqrt{s} \right) }{ d^{3} \Pi'_{3} d^{3} \Pi'_{4} d^{3} \Pi'_{5}
d^{3} \Pi'_{\bar{p}}}  \nonumber \\
&   & (1- f(x,\Pi'_{3})) (1- f(x,\Pi'_{4}))
 (1- f(x,\Pi'_{5})),   \label{pwahr}
\eea
where the quantities $\Pi_{i} \, (i=1,..,5)$ denote the in-medium
momenta of the participating baryons. $\Pi_{\bar{p}}$ and $E_{\bar{p}}$
stand for the \barp \  effective momentum and energy while $s=(\Pi_{1}^{\mu} +
\Pi_{2}^{\mu})^{2}$ is the squared invariant energy available in the
corresponding baryon-baryon collision. An integration over the impact
parameter then yields the Lorentz invariant differential
production cross section
\bea
E_{\bar{p}} \frac{d^{3} \sigma_{\bar{p}}}{d^{3} \Pi_{\bar{p}}}
 =  2 \pi  \int d b  \: b \ E_{\bar{p}} \frac{d^{3} P(b)}{d^{3} \Pi_{\bar{p}}}.
\label{diffwqpr}
\eea
We assume the elementary \apr production cross section $\sigma_{BB\rightarrow
\bar{p}+X} (\sqrt{s})$ in all baryon-baryon channels to be equal to
$\sigma_{pp \rightarrow  \bar{p}+X}(\sqrt{s})$
and employ a parametrization of the free cross section given
by \cite{Batko91} (cf. fig. 3; solid line):
\be
\sigma_{BB \rightarrow \bar{p} +X} = 0.01(\sqrt{s} - \sqrt{s_{0}})^{1.846}
[mb],
\label{proparm}
\ee
with $\sqrt{s_{0}} = 4 m_{N}$ and $ m_{N} = 0.9383 \: GeV/c^{2} $. The
dashed and the dotted line in fig. 3 represent extreme alternative
parametrizations that will also be used below.

While in free space the threshold for the elementary production reaction is
obviously 4 times the nucleon restmass in the medium one additionally
has to take into
account the \ses of all participating particles.
The conservation of energy and momentum has to be guaranteed, i.e.
\be
p_{1}^{\mu} + p_{2}^{\mu} = p_{3}^{\mu} + p_{4}^{\mu} + p_{5}^{\mu} +
  p_{\bar{p}}^{\mu},
\label{prodenim}
\ee
which in terms of effective momenta and \ses
(cf. eqs. (\ref{effp}) and (\ref{effquap})) yields
\bea
\Pi^{\mu}_{1} + U_{v}^{\mu}(\mid \vec{p}_{1}\mid,x) +
\Pi^{\mu}_{2} + U_{v}^{\mu}(\mid \vec{p}_{2}\mid,x)  =
\Pi^{\mu}_{3} + U_{v}^{\mu}(\mid \vec{p}_{3}\mid,x) +
\Pi^{\mu}_{4} + \nonumber \\
U_{v}^{\mu}(\mid \vec{p}_{4}\mid,x)
+\Pi^{\mu}_{5} + U_{v}^{\mu}(\mid \vec{p}_{5}\mid,x) +
\Pi^{\mu}_{\bar{p}} + U_{v}^{\bar{p} \, \mu}(x). \label{gl61}
\eea
With the abbreviation
\bea
\Delta^{\mu}  \equiv & U_{v}^{\mu}(\mid \vec{p}_{3}\mid,x) +
		    U_{v}^{\mu}(\mid \vec{p}_{4}\mid,x) +
		U_{v}^{\mu}(\mid \vec{p}_{5}\mid,x) +
		U_{v}^{\bar{p} \, \mu}(x) \nonumber \\
	   & - U_{v}^{\mu}(\mid \vec{p}_{1}\mid,x)
	   -    U_{v}^{\mu}(\mid \vec{p}_{2}\mid,x)    \label{defde6}
\eea
we obtain in shorthand form
\be
\Pi^{\mu}_{1} + \Pi^{\mu}_{2}  = \Pi^{\mu}_{3}  + \Pi^{\mu}_{4} +
\Pi^{\mu}_{5}  + \Pi^{\mu}_{\bar{p}} + \Delta^{\mu}. \label{gl692}
\ee
When evaluating the threshold for the reaction in terms of effective
momenta, i.e.
\be
\vec{\Pi}_{\bar{p}} + \vec{\Pi}_{3} + \vec{\Pi}_{4} +
	\vec{\Pi}_{5} = \vec{0},
\label{gl65}
\ee
we in general encounter a nonvanishing sum of the vector
\ses of all participating particles. This leads to the
following expression for the threshold
\be
\sqrt{s_{0}} = \sqrt{(m_{\bar{p}}^{\star} + m_{3}^{\star} +
	       m_{4}^{\star} + m_{5}^{\star} + \Delta^{0})^{2}
	      - \vec{\Delta}^{2}}.
\label{thresh}
\ee
In the
$\sigma-\omega$-model, where the \barp-\ses result from
the corresponding baryon \ses by applying  charge conjugation,
the quantity $\Delta^{\mu}$ vanishes and eq. (\ref{thresh}) reduces
to an expression equivalent to the free threshold with the restmasses
replaced by the effective masses of the participating particles.

In order to derive an expression for the differential \barp-multiplicity
we assume, as in refs. \cite{Danielewicz90,Shor90,Batko91}, that the
differential elementary \apr production cross section is
proportional to the phase-space available for the final state:
\bea
E_{3} E_{4} E_{5}E_{\bar{p}} \frac{d^{12}
\sigma_{BB \rightarrow NNN+\bar{p} }
\left(\sqrt{s} \right)}{ d^{3} \Pi_{3} d^{3} \Pi_{4} d^{3} \Pi_{5}
d^{3} \Pi_{\bar{p}}}  =  \sigma_{BB \rightarrow NNN + \bar{p}} (\sqrt{s})
\frac{1}{16 \, R_{4} ( \sqrt{s})}  \nonumber \\
    \delta^{4}
   ( \Pi_{1}^{\mu} +\Pi_{2}^{\mu}- \Pi_{3}^{\mu}
 -\Pi_{4}^{\mu} -\Pi_{5}^{\mu}
 -\Pi_{\bar{p}}^{\mu}-\Delta^{\mu}).
\label{diff12}
\eea
Here, the $\delta$-function guarantees the energy and momentum conservation and
$\sqrt{s}$ is the invariant energy available for the quasi-particles in
the initial state. $R_{4}(\sqrt{s})$ is the 4-body phase-space integral
\cite{Byckling}; it has been included to ensure that the differential
cross section is normalized to the total cross section.

\subsection{Sensitivity to the elementary cross section}

Now we turn first to the analysis of the \apr production mechanism itself and
consequently neglect all in-medium propagation and absorption effects.
For the sake of numerical
simplicity we calculate here the \barp-production using the baryon-\ses
obtained from the nonlinear $\sigma-\omega$-model \cite{Lang} where the
\barp-\ses are determined by charge conjugation from those of the
baryons.

Since we are dealing with subthreshold particle production the
\apr cross section will depend strongly on the behaviour of
the elementary cross section (\ref{proparm}) close to
threshold. Since there are no experimental data available for $\sqrt{s}
- - \sqrt{s_0} \leq 1$ GeV  (cf. fig. 3) we have to rely on an extrapolation
of the data to threshold. In order to investigate the dependence of our
results on the parametrization of the cross section we display in
fig. 4 the invariant differential production \prob for the
reaction $Si+Si$ at 2.1 GeV/u for different extreme parametrizations
 corresponding to the dashed and dotted lines in fig. 3.
This uncertainty of up to one order of magnitude has to be kept in mind when
drawing any conclusions from the comparision of our work with the experimental
data. For our further analysis we will adopt the parametrization
(\ref{proparm}).\

Due to these uncertainties it is justified to neglect the Pauli-blocking
effects in the calculation of \apr production which turn out to
reduce the production \prob by about 10 to 20\% as shown in fig. 5
for the special case of
a central $Au+Au$ collision at 2.1 GeV/u.
The effects of the Pauli-blocking increase with decreasing beam
energy and become less important for lighter systems as e.g. $Ne + Ne$.

\subsection{Probing the high density phase with \barp}

In order to demonstrate the sensitivity of the \apr production to the
equation of state (EOS) for nuclear matter we display in fig. 6 the invariant
differential \barp-production \prob for the reactions
$Au+Au$ and $Si+Si$ at 2.1 GeV/u calculated for different parametrizations
of the EOS. The parametersets NL1 and NL3 (cf. ref. \cite{Lang})
employ the same incompressibility
$K = 400$ MeV, but differ in the effective nucleon mass at $\rho = \rho_{0}$
($m^{\star}/m = 0.83$ for NL1; $m^{\star}/m = 0.7 $ for NL3), while
NL1 and NL2 employ the same effective mass, but differ in their
incompressibility ($K = 200 $ MeV for NL2). For both systems we observe
a production \prob which is enhanced by roughly one order of
magnitude for NL3 as compared to that from NL1 and NL2.
The reason for this behaviour is obviously the low lying threshold which
goes along with the small effective nucleon mass when using the
prametrization NL3.

While there is hardly any sensitivity to
the incompressibility (NL1 versus NL2) in the light system, one can
observe an increase of the production \prob by a factor of 3
with decreasing incompressibility in the heavy system.
The reason for this sensitivity is given in fig. 7 where
we show the differential \barp-multiplicity $dN/d\rho$ as a function
of the baryon density $\rho/\rho_{0}$ at the individual \apr production
point for $Si+Si$ (dashed line; multiplied by a factor of 50) and
$Au+Au$ at 2.1 GeV/u. It is clearly seen that almost all  \aps are produced
at $\rho \ge 3\rho_{0}$ in the heavy system while the majority of \aps is
created between $2 \rho_{0}$ and $3\rho_{0}$ in the light system. This
difference is easily understood in terms of the pile-up of density in
the heavy system. In the case of $Au + Au$ reactions high densities are
obtained for parametersets that describe a rather soft EOS \cite{Lang92}.
Thus compared to NL1 the parameterset NL2 leads to a
larger density in the
reaction zone resulting also in a higher \apr production cross section.

\subsection{Variation with the impact parameter}

Since central collisions lead only to minor contibutions
to the inclusive cross section in \hics it is of interest to study the
\barp-multiplicity as a function of the impact parameter b.
For this purpose we show in fig. 8 the calculated differential \apr
multiplicity for $Si+Si$ at 2.1 GeV/u multiplied by $2\pi b$ as a function
of b (solid line). This quantity can be approximately fitted by (dotted line)
\be
\frac{d \sigma (b)}{db} = 2 \pi b P(b=0) \ S(b),
\label{sizussvonb}
\ee
where S(b) represents the geometrical overlap of two nuclei with mass number
$A_{1}$ and $A_{2}$ as a function of the impact parameter. Assuming
$A_{1} > A_{2}$ the overlap function S(b) reads
\bea
S(b)  = \frac{1}{\pi R_{2}^{2}} \left\{ R_{1} \arccos \left[
\frac{b^{2} +R_{1}^{2} -R_{2}^{2}}{2 R_{1} b} \right] +
 R_{2} \arccos \left[
\frac{b^{2} +R_{2}^{2} -R_{1}^{2}}{2 R_{2} b} \right] \right.\nonumber \\
 \left. - \frac{1}{4b^{2}} \left[ (b^{2} + R_{1}^{2} - R_{2}^{2}) +
      ( b^{2} + R_{2}^{2} - R_{1}^{2}) \right]
   \sqrt{2(R_{1}^{2}+R_{2}^{2})\:b^{2}
      - b^{4} - (R_{1}^{2}-R_{2}^{2})^{2}} \right\} \nonumber \\
\mbox{for } R_{2} \leq b \leq R_{1}   \nonumber
\eea
and
\bea
S(b)  =  1  \nonumber \\
\mbox{for } b < R_{2} < R_{1} \label{svonb}
\eea
with $ R_{1} = a A_1^{\frac{1}{3}}$ and $ R_{2} = a A_2^{\frac{1}{3}}$.
In case of symmetric systems this reduces to
\be
S(b) = \frac{1}{\pi R^{2}} \left\{ 2 R \arccos \frac{b^{2}}{2Rb} -
b \left( R^{2} -\frac{b^{4}}{4} \right)^{\frac{1}{2}} \right\}.
\label{ssysmm}
\ee
A more systematic analysis with respect to $A_1$ and $A_2$ shows that an
optimal fit to the impact parameter dependence is obtained for the
radius parameter $a = 1.1 fm$.

\subsection{Baryonic decomposition}

It is well known from transport
calculations \cite{Batko91,Lang92} that heavy mesons and \aps
are dominantly produced via multiple baryon-baryon collisions where
especially nucleon-resonance channels play an important role.
To gain further insight in
the mechanism of \barp-production we analyse the relative contribution
from various channels in more detail by counting the number of
collisions $ N_{1}+N_{2} $ that two baryons have undergone before
producing an \apr in a mutual collision and determine the corresponding
production \prob as a function of $ N_{1}+N_{2} $ for the channels
$NN, \, N\Delta$ and $\Delta\Delta$ separately. The resulting distributions
are displayed in fig. 9 for the reactions $Si+Si$, $Ca+Ca$ and
$Au+Au$ at 2.0 and 2.5 GeV/u. The full histograms denote the channel
$NN \rightarrow\bar{p}+X$, the open histograms denote the channel
$N\Delta\rightarrow\bar{p}+X$ and the hatched histograms the
$\Delta\Delta\rightarrow\bar{p}+X$, respectively. Table \ref{tab1}
shows the relative contributions
integrated over $N_{1}+N_{2}$. From table 1 we can extract a general trend:
The $NN$-channel is obviously of minor importance.
In heavier systems the contributions from resonances to
the \barp-production are generally higher as compared to lighter systems.
This correlates with the higher density pile-up in heavy systems.
The resonances become, furthermore, more important with decreasing
beam energy.

In order to study this phenomenon in more detail
we display in fig. 10 the relative contributions to the \barp-yield
for $Ca+Ca$ as a function of the beam energy.
The contribution from the $NN$-channel is below
the 10\% level for $E_{beam} \leq 2.5 GeV/u$.
 While for higher energies the $\Delta N$-channel
dominates the production of \aps the $\Delta\Delta$- and the $\Delta N$-
channel are approximately equally important in the energy regime between
2.4 GeV/u and 2.0 GeV/u. Below 2.0 GeV/u the $\Delta \Delta$-channel
is clearly the dominant one.

{}From the analysis of this section we can conclude that, as far as
\apr production is concerned, the nucleon-resonances act as
energy reservoirs that can release their energy in baryon-baryon collisions
to allow for \barp-production. That the resonances are populated in
sufficiently large numbers is a consequence of the formation of
resonance matter \cite{Ehehalt} in relativistic heavy-ion collisions.

\subsection{Sensitivity to in-medium \barp-properties}

The threshold for the elementary \barp-production reaction
according to (\ref{thresh})  depends on the in-medium properties
of the \aps. In order to investigate this dependence we calculate the
\barp-production \prob for $Ni+Ni$ at 1.85 GeV/u (central collision) for
a variety of different \apr \ses: In the calculation shown in the
upper part of fig. 11 the \barp-production \prob was determined
using a fixed vector part of the \apr \se ($U_{0}^{\bar{p}} = -98$ MeV
at $\rho = \rho_{0}$) while varying the scalar part $U_{s}^{\bar{p}}$
(see figure caption). The results displayed in the bottom
part of fig. 11 were obtained by varying the vector part while keeping
the scalar part $U_{s}^{\bar{p}}$  fixed at $-159$ MeV at $\rho =
\rho_{0}$. The same analysis performed for other systems than
$Ni+Ni$ lead to quite similar results. Thus we conclude that the
\apr production cross section exhibits an extreme sensitivity to the
in-medium \barp-properties. This sensitivity will be used below to obtain
approximate values for the $\bar{p}$ selfenergy in comparison to
experimental data.

\section{In-medium $\bar{p}$ propagation and absorption}
\label{apabs}
The \aps produced in individual baryon-baryon collisions can
be annihilated while propagating out of the dense nuclear medium.
Since we expect the \barp-annhilation to dominate the
\apr spectra a proper treatment of this reaction channel is crucial for
a comparison with experimental data. In order to be able to treat the
\barp-absorption and propagation sufficiently well we employ a
new numerical method which allows to calculate the propagation as well as
the \apr absorption nonperturbatively. We neglect the elastic
scattering of antiprotons with the surrounding baryons; according to
ref. \cite{Ko93} this elastic scattering only causes a flattening of
the \apr momentum spectra. The numerical details of our
method are presented in Appendix A. In this section we concentrate on the
effects of in-medium absorption and propagation of the \aps.

During a \hic or a \pnr the \aps are created with effective masses
$m^{\star }_{\bar{p}}$ and momenta $\Pi_{\bar{p}}^{\mu}$. These
particle properties change according to the surrounding medium as the
\aps leave the reaction zone. At the end one observes free \aps
with kanonical momenta $p^{\mu}_{\bar{p}}$. Since the properties
of the \aps are determined by their optical potential one expects
significant effects of the \barp-propagation on the \barp-spectra.
We analyse these effects in figs. 12 and
13. Fig. 12 shows the variations in the \barp-spectra
due to different \apr \ses for a central $Si+Si$ collision at 2.1 GeV/u
at $\Theta_{lab} = 0^{o}$. The production \prob is identical for all
cases. The calculations shown in the top part of this figure
were performed for $U_s^{\bar{p}} = 0 $ and
different vector parts (see fig. 12). The bottom part of this figure
shows the result of calculations for $U_{\mu}^{\bar{p}} = 0$ and
varying scalar parts of the \apr \se. Comparing both figures
we conclude, as for the \barp-production itself, that the effects of
the scalar and vector part of the \se on the \barp-spectra are
qualitatively and quantitatively similar. Attractive potentials result
in a reduction of the \barp-momenta as the \aps leave the reaction zone
while repulsive potentials lead to enhanced \barp-momenta.

Fig. 13 displays the same analysis for $p+Cu$ at 4.0 GeV
at $\Theta_{lab} = 0^{o}$. Again, the production was calculated
identically for all three cases. Comparing the \barp-spectrum obtained
for free propagation ($U_{sep}=0$) with the one obtained using a
weakly attractive optical potential ($U_{sep}= -90$ MeV) one observes a
slight shift of the spectrum towards lower momenta. This shift is
easily explained by the loss of kinetic energy that occurs as the
effective mass $m_{\bar{p}}^{\star}$ increases to the value of the
restmass when the \aps leave the reaction zone. Using a strongly attractive
optical
potential ($U_{sep}=-560$ MeV) one realizes in addition to this shift  a
reduction of the spectrum due to the propagation effects. Contrary
to  \hics the target-nuclei in \pncs are not destroyed during those
timescales when the antiprotons move into the continuum. This means
that the target nuclei build a potential barrier  which
the \aps have to overcome when leaving the nuclei. The low
energy \aps cannot overcome this potential wall and get trapped in
the nucleus which leads to the observed reduction in the \barp-spectrum.

Figs. 12 and 13 clearly show that the \apr in-medium
propagation has an impact on the differential \barp-spectra.
Therefore it is important to apply the complete mean-field dynamics of
the RBUU model when describing the propagation of \aps.

Now we turn to the in-medium \apr absorption ($\bar{p}+B\rightarrow X$).
Due to the fact that there is no information available on the in-medium
cross section we have to rely on the free cross section in the
parametrization from \cite{Koch89}(6).
 In our transport calculation we replace the
free $s-s_{0}$ with that determined from the in-medium properties
of the colliding particles, i.e.
$s = (\Pi^{0}_{\bar{p}}+\Pi^{0}_{B})^{2}-(\vec{\Pi}_{\bar{p}}+
\vec{\Pi}_{B})^{2}$ is the squared invariant energy available in
the elementary antiproton-baryon collision while
$s_{0}=(m^{\star}_{\bar{p}} + m_{B}^{\star})^{2}$ denotes the squared sum of
the effective masses of the colliding particles. The dashed line in
fig. 1 represents a cutoff at 100 mb which is  introduced
to simulate possible in-medium screening effects. The value
of 100 mb for the cutoff is in line with the annihilation radius for
proton-antiproton annihilation derived in a optical model calculation
\cite{Weise93}. Fig. 14 shows the number of antiproton-baryon collisions
as a function of the corresponding value of $\sigma_{abs}$ for a central
reaction $Si+Si$ at 2.1 GeV/u. Since only a small fraction
of all events lies in the region of $100 - 192$ mb our results do not
depend significantly on the cutoff. Similar studies for \pncs
show even less sensitivity to this cutoff since the antiprotons move with
higher momenta with respect to the target nucleus.

In order to study the \barp-absorption effects on the \apr spectra
from \hics we have performed calculations using different fixed absorption
cross sections in comparison to the dynamically determined cross section
(\ref{para}). Fig. 15 shows a full calculation for the
system $Si+Si$ at 2.1 GeV/u at $\Theta_{CM}=0^{o}$. The top
line denotes the calculated \barp-\prob without absorption
 using the baryon \ses obtained from the nonlinear
$\sigma-\omega-$model with \barp-\ses obtained via charge
conjugation from the baryon selfenergies. The lower lines represent
calculations
with different constant absorption cross sections while the solid line results
 from a calculation
using the parametrization (\ref{para}).
This figure clearly reflects
the dominant role of the in-medium absorption of \aps which
leads to a reduction in \barp-\prob by approximately two orders of
magnitude; these absorption rates agree with those determined by Li et al.
\cite{Ko93}. Furthermore, the spectral distribution obtained
when using the parametrization (\ref{para}) is - except for small
deviations at low \apr momenta - almost identical to the distribution
obtained for a constant cross section of 70 mb. This value
corresponds approximately to the mean value of the distribution
displayed in fig. 14.

In this context we note that in \pnrs the corresponding mean value of the
elementary \barp-absortion cross section is aproximately 50 mb which
leads to absorption factors of $10 - 30$
depending on the size of the target nucleus and the beam energy.
The difference in \apr absorption between \hics and \pnrs is mainly due
to the different kinematics in both cases: During a \hic
most of the
\aps are created in the CMS of the collision which implies that
the relative velocity of the \aps and the surrounding baryonic medium
is low. This leads to a small $s-s_{0}$  for the annihilation
reaction and results in high values of the absorption cross section
(see fig. 1). In contrast to this situation the \aps
produced in \pncs move with momenta of around 1 GeV/c through the
baryonic medium which leads to smaller annihilation rates.
Another difference in both types of reactions is that the \barp-\abs
in \pncs takes place at a maximum density $\rho_{0}$ while the
\aps produced in \hics experience higher densities (cf. fig. 7)
which again leads to an increase in the absorption probability.

Finally we study the impact parameter dependence of the \apr absorption.
For this purpose we have to investigate a physical quantity
that does not show the impact parameter dependence of the \apr
production mechanism. The ratio $R$ of the differential \barp-\prob
calculated including \abs and the differential production \prob
meets this requirement. In Fig. 16 we display $R$ for
$Ni+Ni$ at 1.85 GeV/u, $Ne+Ni$ at 2.0 GeV/u and $Si+Si$ at 2.1 GeV/u
as a function of $b/b_{max}$ where b denotes the impact parameter and
$b_{max}$ is the sum of the corresponding radii of target and
projectile. The ratio $R$ is evaluated for zero \apr momentum in the CMS of the
corresponding \hic. The dots represent the values of $R$ calculated
in the RBUU model and the solid lines correspond to fits to these
numerical data using the function
\be
R \left( b/b_{max} \right) = A - (1 +
	    e^{ \frac{ \left( b/b_{max} - B \right) }{C}  })^{-1}.
\label{fit5}
\ee
The parameters A, B and C are given in table \ref{tab2}.
Except for peripheral collisions ($b/b_{max} > 0.6$) the ratio R is determined
by the constant A for all systems. This means that the \barp-production
spectra for $b/b_{max} \le 0.6$ are reduced due to the absorption
mechanism by a constant factor determined by the size of the system.
The parameters B and C are similar for all systems considered. This
leads to the conclusion that the absorption mechanism is similar
for all systems and can be understood by means of simple geometrical
considerations.

\section{Comparison with experimental data}

\label{res}
After the systematical analysis of the in-medium \apr production
and absorption mechanism we now turn to the comparison of our
calculations with the most recent data from KEK and GSI. In order to
perform this comparison we describe the \barp-\ses on the basis of the
$\sigma-\omega-$model with free coupling parameters $g^{\bar{p}}_{s}$ and
$g^{\bar{p}}_{v}$. The comparison with the experimental data will
allow to approximately determine these parameters and then provide first
information on the \apr potential in the dense medium.
The quasi-particle properties, i.e. the nucleon selfenergies
$U_s(x, p) ,  U_{\mu}(x, p)$, of the baryons
participating in the \barp-production reaction are taken
from refs. \cite{KLW1,KLW2,Tomo}. $U_s(x,\,p)$ and $U_\mu(x,\,p)$
are fixed to reproduce the saturation properties of nuclear matter,
the empirical proton-nucleus optical potential as well as the density
dependence of $U_s$ and $U_{\mu}$ from Dirac-Brueckner theory \cite{Mal}.
For orientation the actual values
for $U_s(p)$ and the zero'th component of the vector field $U_0(p)$ are
displayed in fig. 17 for $\rho_0 (\approx 0.17 fm^{-3}),
2 \rho_0,$ and $ 3 \rho_0$.

Before we present the results of the calculations using
these expressions for the baryon \ses we give a brief description of the
numerical method used to implement these \ses into the \apr
production process.
Since we deal with explicit momentum dependent \ses ($U_s(x,\,p),\,
U_{\mu}(x,\, p)$) for the baryons their effective momenta and
masses are given by
\bea
\Pi^{\mu}_{j} & = & p^{\mu}_{j} - U_{v}^{\mu}(\mid \vec{p}_{j}  \mid, x)
\label{erg11} \\
m^{\star}_{j} & = & m + U_{s}(\mid \vec{p}_{j} \mid, x) \label{erg21} \\
  j & = & 1,\, \cdots,\, 5,  \nonumber
\eea
while the corresponding quantities for the \aps read
\bea
\Pi^{\mu}_{\bar{p}} & = & p^{\mu}_{\bar{p}} - U_{v}^{\bar{p} \, \mu}(x)
\label{erg13} \\
m^{\star}_{\bar{p}} & = & m + U_{s}^{\bar{p}}(x). \label{erg22}
\eea
The elementary \barp-production events occuring at location $\vec{x}$
are evaluated in the corresponding local rest frame (LRF) of the
nuclear matter ($j_{\mu} = (\rho,0,0,0$)). In this frame the spatial components
of the
vector \ses vanish by definition. This implies that the vector components
of the effective and the canonical momenta of all particles are the same
(\ref{erg11}, \ref{erg13}). The spatial components of the quantity
$\Delta^{\mu}$ defined in eq. (\ref{defde6}) also vanish. Energy and
momentum conservation then yields
\bea
\Pi^{0}_{1} + \Pi^{0}_{2} & = & \Pi^{0}_{3}  + \Pi^{0}_{4} +
\Pi^{0}_{5}  + \Pi^{0}_{\bar{p}} + \Delta^{0}, \label{gl691} \\
\vec{p}_{1} + \vec{p}_{2} & = & \vec{p}_{3} + \vec{p}_{4} + \vec{p}_{5}
+ \vec{p}_{\bar{p}}. \label{gl690}
\eea

In order to calculate the \barp-production \prob we first have to
evaluate the threshold $\sqrt{s_{0}}$ in the LRF. In the CMS of the
particles in the final state
 the threshold is obtained for the kinematical situation
with all particles at rest. Thus in the LRF these particles move
at threshold with identical velocities. The momenta of the baryons and
the \apr do not necessarily have to be equal because the effective masses
of the nucleons and the \apr can differ from each other. In order to
guarantee momentum conservation we use the following ansatz for the
effective masses of the baryons and the \aps
\bea
\mid  \vec{p}_{j}  \mid & = & \frac{m^{\star}(\mid \vec{p}_{j}\mid)}
{3 m^{\star}(\mid \vec{p}_{j}\mid) + m^{\star}_{\bar{p}}}
\mid  \vec{p}_{1} + \vec{p}_{2} \mid,
\label{impan16} \\
\mid  \vec{p}_{\bar{p}}  \mid & = & \frac{m^{\star}_{\bar{p}}}
{3 m^{\star}(\mid \vec{p}_{j}\mid) + m^{\star}_{ \bar{p} } }
 \mid \vec{p}_{1} + \vec{p}_{2}  \mid. \label{impan26}
\eea
Since each baryon mass depends itself on the momentum of the corresponding
baryon these equations are iterative equations for the absolute value of the
momenta. Using these effective masses the threshold for \apr production
in the LRF reads
\be
\sqrt{s_{0}} = \sqrt{(3 \Pi^{0}_{j} + \Pi_{\bar{p}}^{0} + \Delta^{0})^{2}
	- (3  \vec{\Pi}_{j} + \vec{\Pi}_{\bar{p}})^{2} }. \label{so6}
\ee
\barp-production takes place if the invariant energy $\sqrt{s}$
($s = (\Pi_{1}^{0}+ \Pi_{2}^{0})^{2} - (\vec{\Pi}_{1}^{2} +
\vec{\Pi}_{2}^{0})^{2}$) of the baryons in the initial channel
lies above the threshold $\sqrt{s_{0}}$. According to definition
(\ref{diff12}) the elementary \apr production \crs is a function of
the invariant energy $\sqrt{s'}$ available for the particles in the
final state of the production reaction
\be
\sqrt{s'}= \sqrt{ (\Pi_{3}^{0} + \Pi_{4}^{0} + \Pi_{5}^{0} +
\Pi_{\bar{p}}^{0})^{2} - (\vec{p}_{3} + \vec{p}_{4} + \vec{p}_{5} +
\vec{p}_{\bar{p}})^{2}}
\label{gl693}
\ee
Using (\ref{gl691}) and (\ref{gl690}) this reads
\be
\sqrt{s'} = \sqrt{ (\Pi^{0}_{1} + \Pi^{0}_{2} - \Delta^{0})^{2} -
		   (\vec{p}_{1} + \vec{p}_{2})^{2} }. \label{spri6}
\ee
This equation shows that in order to calculate the production \prob one
has to know the momenta of the particles in the final state of the
elementary production reaction when employing momentum dependent
potentials $U_s(x,\,p)$ and $U_{\mu}(x,\,p)$ for the baryons.

Since we are dealing with subthreshold particle production the main
contribution to the production \crs will arise from events with
collinear baryon momenta because this kinematical situation
minimizes the kinetic energy of the baryons. Based on this argument
we assume for the momenta of the baryons in the final state
\be
\vec{p}_{j} = \frac{1}{3} ( \vec{p}_{1} + \vec{p}_{2} - \vec{p}_{\bar{p}} ).
\label{pdef634}
\ee

We have applied the above mentioned formalism to evaluate the antiproton
cross section for the reactions $p + ^{12}C$ and $p + ^{63}$Cu at
bombarding energies of 5, 4,  and 3.5 GeV. The
corresponding invariant cross sections in comparison with the data of
ref.  \cite{KEK} are shown in fig. 18 as a function of
the momentum of the emitted antiproton in the lab. frame at $\Theta = 0^o$,
assuming free antiprotons, i.e. $g_s^{\bar{p}}, g_v^{\bar{p}} = 0$.
The calculations slightly underestimate the experimental data, but
already approximately reproduce the shape of the momentum-spectra as well as
the dependence on bombarding energy and mass. The
$\bar{p}$ reabsorption amounts to a factor of 12 in case of $^{12}C$ and to
a factor of 19 for $^{63}Cu$ roughly in line with simple geometrical estimates.

When adjusting the constant
$g_s^{\bar{p}}$ such that the scalar potential (\ref{e660}) becomes slightly
attractive ( $- 50$ to $- 100$ MeV at $\rho_0$) the reproduction
of the data improves at all energies significantly which is exemplified for
4.0 GeV by the dashed line in fig. 18. In the above comparison
we cannot distinguish between scalar and vector antiproton selfenergies because
both yield similar results for the $\bar{p}$ spectrum if the same
Schr\"odinger-equivalent optical potential is achieved.
Furthermore, when using antiproton selfenergies in line with the relativistic
mean-field theories \cite{Serot}, i.e. changing only the sign of
the nucleon vector potential, we overestimate the $\bar{p}$ yield by more than
an order of magnitude at all energies for both systems.

We now turn to the nucleus-nucleus case. The calculated antiproton invariant
differential cross section for the reaction $^{28}Si+^{28}Si$ at 2.1 GeV/A
and Ni + Ni  at 1.85 GeV/u is shown in fig. 19 in comparison
to the experimental data of ref.\cite{BEVALAC1} and ref. \cite{GSI};
the upper lines represent the results of  the calculations for
free antiprotons without including any reabsorption. When taking care of
antiproton annihilation according to eq. (\ref{eq:ann}) the
yields drop to the lower
full lines. In the case of $Ni+Ni$ the data are now underestimated sizeably.
However, using
attractive scalar (or vector) selfenergies at $\rho = \rho_0$ of
about $- 100$ to $-150$ MeV we reproduce the data for $Ni+Ni$, for
$Si+Si$, however, we miss the data point at $p = 1$ GeV/c.

We thus use a potential of $-100$ MeV at $\rho = \rho_{0}$
to predict the differential \barp-excitation function in $Ni + Ni$
collisions from 1.4 to 2.5 GeV/u (fig. 20), a system that will be
explored at GSI in the near future \cite{kienle}.

The different
value for the attractive antiproton field at $\rho = \rho_0$
in $p + A$ and $A + A$ reactions is
due to the fact that in $p + A$ collisions the antiprotons move with
momenta of 1 - 2 GeV/c with respect to the nuclear medium, whereas in
$A + A$ collisions the antiprotons have smaller momenta in the nucleus-nucleus
center-of-mass frame. In view of uncertainties of our present studies with
respect to the elementary production
cross sections close to the thresholds we provide
areas for the antiproton Schr\"odinger equivalent potential at $\rho = \rho_0$
in fig. 21, as extracted from the comparison with the experimental
data for $p + A$ and $A + A$ reactions. These areas are far from the values
expected by charge conjugation from the familiar $\sigma - \omega$ model
\cite{Serot} (dashed line) and thus exclude
relativistic mean-field models with the
same parameter-sets for nucleons and antinucleons. However, our extracted
values are well in line with a Schr\"odinger-equivalent potential (solid line
in fig. 21) as calculated from the dispersion relation
(see equations (\ref{dispses1}) - (\ref{cass5})). Crucial for this result is
the correct momentum dependence for the $p+A$ potential in the entrance
channel. If we use the original Walecka model without the momentum
dependent coupling strength we are forced to compensate for the strong
repulsion with much deeper \apr potentials.

The primordial \apr production rates and the reduction due to absorption
obtained here agree well with those obtained by Li et al. \cite{Ko93}.
These authors also get agreement with the data for $Si+Si$, however,
by using the
very deep \barp-potential obtained by charge conjugation from the
Walecka model and the corresponding strong repulsion in the
entrance channel; using the same model we also reproduce the data, as well
as their calculations within a factor of two.
However, the entrance channel potential is much too repulsive,
as mentioned in section 3.1, so that an unphysically deep
antiproton potential is needed to compensate for this repulsion.
This fact shows that the conclusion of ref. \cite{Ko93} that
the agreement of their calculation with the data supports the assumption
of a very deep G-parity transformed nucleon potential for
the \aps is not justified.

\section{Summary}

In this work we have evaluated the differential cross section for $\bar{p}$
production for proton-nucleus and nucleus-nucleus reactions in the
subtreshold regime by considering on-shell baryon-baryon production channels
involving nucleons and $\Delta$'s with their in-medium quasi-particle
properties and treating $\bar{p}$ propagation and annihilation
nonperturbatively. The quasi-particle properties of the nucleons are fixed
in our approach by the nuclear saturation properties, the proton-nucleus
empirical potential as well as Dirac-Brueckner calculations at higher
density. By varying especially the antiproton selfenergy we have shown that
the differential $\bar{p}$ spectra are highly sensitive to the antiproton
quasi-particle properties. Though we still have to rely on proper
extrapolations to threshold of the elementary process $p + p \rightarrow
\bar{p} + X$, the latter sensitivity can be used to obtain approximate
values for the $\bar{p}$ Schr\"odinger-equivalent potential from a systematic
comparison to the available experimental data.

In this respect we have performed a systematic study of p-nucleus and
nucleus-nucleus collisions in a broad kinematical regime and compared our
numerical results to the data from KEK \cite{KEK} and GSI \cite{GSI}.
We find a consistent description of all data employing a rather weak
attractive potential for the antiprotons which is well in line with a
dispersive potential extracted from the dominant imaginary part of the
antiproton selfenergy due to annihilation. Essential for this result is the
use of the correct momentum dependence of the nucleon-nucleus potential
in the entrance channel. Although the validity of the simple dispersive model
used here may be questionable at higher densities and the extrapolation of
the elementary \barp-production cross section down to threshold
introduces an uncertainty, the consistency of the \barp-potentials
obtained from a fit to the data with those calculated in the dispersive
model indicates that the basic \apr production mechanism in
$p+A$ and $A+A$ reactions is understood.

It has become clear that the baryonic production channels at
subthreshold energies involve dominantly one or two nucleon resonances
and that the production
proceeds at the highest densities that can be reached in a nucleus-nucleus
reaction. Only at these high densities the relative population of nucleon
resonances ($\approx$ 30 \% as shown in \cite{Ehehalt}) as well as the
nucleon-resonance collision rate are high enough to allow for $\bar{p}$
production. Otherwise the resonances decay to a nucleon and a pion before
colliding with another baryon such that the energy stored in the resonance
gets lost for the production.

We note in closing that the antiproton production studies at the AGS
\cite{Koch,AGS1,AGS2} around 15 GeV/u, although far above the free production
threshold, might yield further information on the dynamics and selfenergies
of antiprotons at even higher densities.

\vspace{1cm}
\noindent
The authors acknowlegde valuable discussions with A. Gillitzer, P. Kienle,
W. K\"onig and A. Schr\"oter as well as their information on experimental
results prior to publication. They are also very grateful to C. M. Ko
for detailed discussions both on the physics and on the computational
aspects of these calculations.

\newpage

\begin{appendix}
\section{The weighted testparticle method}
In the following we present the weighted testparticle method used
 to implement the \apr production,
absorption and propagation in the RBUU model. It is especially
suited for the simulation of processes with low production
probabilities and/or high in-medium absorption \crss for which
conventional testparticle methods fail because of low statistics.

Assuming that during a calculation N elementary baryon-baryon
collisions with sufficient energy for \apr production occur, we define for
each collison $i$ ($i \epsilon \{1,\  \cdots,\ ,N \}$) a three-dimensional
momentum grid in the reference frame (CMS of the \hic or lab-system).
In order to calculate the production \prob for each collision $i$ we
transform this grid into the system that fullfills eq. (\ref{gl65}) or
into the LRF (see eqn. (\ref{erg11}) to (\ref{pdef634})). After
calculating the invariant differential \barp-production \prob
$ w^{(i)}_{klm}$ for each grid point $ \vec{p} = (k \Delta p_{x}, \,
l \Delta p_{y}, \,  m \Delta p_{z} ) $ the momentum grid is transformed
back into the reference frame. To obtain the total invariant
differential production \prob without absorption
for a \hic or a \pnr one has to sum up the
contributions from all the N elementary production reactions
\be
      W_{klm}           = \sum_{i=1}^{N} w^{(i)}_{klm}.
\label{gttmprwahr}
\ee
In order to calculate the \apr \prob including propagation through the
dense medium and absoption we start out with the momentum grid described
above. For all N baryon-baryon collisions contributing to the
\apr production we place a testparticle representing an \apr on every
point of the momentum grid. This yields additional j ($j = k \cdot l\cdot m $)
testparticles for each of the N baryon-baryon collison events. The effective
masses and momenta of these testparticles are determined according to
both the location on the momentum grid and the spatial coordinate
where the baryon-baryon collision takes place. In addition we attach
variable weights to the testparticles. The weight $w'^{(i)}_{h}$ of the
h-th testparticles 'created' in the i-th collision denotes the
invariant differential production \prob $w^{(i)}_{klm}$ at the
corresponding grid point. The result of this procedure is an
ensemble of testparticles representing the produced \aps that contain
information about the differential \barp-distribution for each
relevant baryon-baryon collision. The entire testparticle
distribution together with the corresponding weights represents the
phase-space distribution of the \aps produced in a \hic or a
proton-nucleus reaction.

We now describe how we treat this ensemble of testparticles in order to
calculate the \apr propagation and absorption: Each testparticle is
propagated by means of the equation of motion for the antiparticle
phase-space distribution function (\ref{vlat}).
At the end of each timestep the absorption rate
for the testparticles is calculated according to their weights. In order
to perform this task we convert the collision integral (\ref{collabs32})
into a differential equation for the  testparticle weights
\bea
\frac{d w'^{(i)}_{h} }{dt} \mid_{x,\Pi_{\bar{p}}} & = & - \frac{4}{(2\pi)^{3}}
\int \int \frac{d^{3}\Pi_{B}}{\Pi_{B}^{0}} d^{4}\Pi_{X} m_{B}^{\star}
W(\bar{p} + B \rightarrow X) \nonumber \\
&   &\delta^{4}(\Pi_{\bar{p}}^{\mu}
+\Pi_{B}^{\mu} - \Pi_{X}^{\mu})  f(x^{\mu},\Pi_{B}^{\mu}) w'^{(i)}_{h}\mid_{
x,\Pi_{\bar{p}}},
\label{eq:ann}
\eea
where $x^{\mu}$ is the space-time coordinate where the
annihilation takes place. $\Pi_{\bar{p}}, \,
\Pi_{B}, \, m^{\star}_{\bar{p}}$ and $m^{\star}_{B}$ denote the effective
momenta and masses of the \aps and the baryons, respectively. This
differential equation leads to the following changes in the testparticle
weights in timesteps $\Delta t$,
\bea
w'(t_{j+1})^{(i)}_{h}  & = & w'(t_{j} + \Delta t)^{(i)}_{h} \nonumber \\
& = &  w'(t_{j})^{(i)}_{h} \, exp(-\int_{t_{j}}^{t_{j+1}} \int
\frac{d^{3}\Pi_{B}}{\Pi_{B}^{0}}  m_{B}^{\star} W(\bar{p} + B \rightarrow X)
f(x^{\mu},\Pi^{\mu}_{B}) dt).
\label{tests}
\eea
Solving the integral in the exponent of (\ref{tests}) by means of
the local density approximation (LDA) \cite{Lang93} leads to
\be
w'(t_{j+1})^{(i)}_{h} = w'(t_{j})^{(i)}_{h} exp( -
   \frac{1}{\Delta V N_t} \sum_{j \varepsilon V_{i} }
 \mid \,   \frac{  \vec{\Pi}_{\bar{p}} }{\Pi^{0}_{\bar{p}} } -
  \frac{ \vec{\Pi}_{B_{j}}  }{\Pi^{0}_{B_{j}} }\, \mid
  \sigma (\bar{p}+B_{j} \rightarrow X ) \Delta t ),
\label{ldnsum}
\ee
with $N_t$ being the number of testparticles per nucleon used to model
the baryon phase-space distribution and $\Delta V$ being the normalization
volume used in the RBUU model to evaluate densities and currents.
Eq. (\ref{ldnsum}) allows for a nonperturbative evaluation of
high absorption rates without a reduction of the number of testparticles.

At the end of the simulation we project the weight of each \apr
testparticle obtained in the final timestep $t_{f}$ of our calculation
on the three-dimensional momentum grid (see above)
\be
 \tilde{w'}^{(i)}_{h} (t_{f},\, p_{x},\,p_{y},\,p_{z})
\rightarrow  \tilde{w'}^{(i)}_{klm}
\label{gttmgl4}
\ee
and sum up all contributions in order to obtain the invariant differential
\apr probability for the corresponding \hic or \pnr.

\newpage
\section{Figure Captions}
\newcounter{figno}
\begin{list}
{\underline{fig.\arabic{figno}}:}{\usecounter{figno}
	 \setlength{\rightmargin}{\leftmargin}}

\item    Free annihilation cross section for $\bar{p}+p \rightarrow X$.
         Dots: experimental data from
	 \protect\cite{annihilationdata}; solid line: parametrization
	 (\protect\ref{para}).

\item        Real part of the \apr \se as a function of its kinetic energy
using
         different parameters B in (\protect{\ref{para}})
         for the annihilation cross section.

\item        Inclusive antiproton cross section in a pp collision. The
	 solid line indicates the parametrization (\protect\ref{proparm}).
	 The dashed and dotted lines show 'extreme' alternative
        extrapolations to the
         threshold. The experimental data are taken from
	 \protect\cite{Danielewicz90}.

\item    Invariant differential production probability for $Si+Si$ at
         2.1 GeV/u in the CMS ($\Theta=0^{o}$; b = 0; EOS: K = 200 MeV,
	 $m^{\star}/m = 0.7$ at $\rho = \rho_{0}$) using different
	 parametrizations for the elementary production cross section
	 according to fig. 3.

\item    Effects of Pauli-blocking for $Au+Au$ at 2.1 GeV/u in the
	 CMS ($\Theta = 0^{o}, \, b = 0$); Solid line: Pauli-blocking
	 included; dashed line: no Pauli-blocking.

\item    Differential \barp-production probability for $Si+Si$ (top)
         and$Au+Au$ (bottom) at 2.1 GeV/u in the CMS ($\Theta=0^{o}$,
	 $b=0$) using different parametrizations for the EOS;
	 NL1: K = 400 MeV, $m^{\star}/m$ = 0.83; NL2: K = 200 MeV,
	 $m^{\star}/m$ = 0.83; NL3: K = 200 MeV, $m^{\star}/m$ =
	 0.83 at $\rho = \rho_{0}$.

\item    Production probability as a function of $\rho/\rho_{0}$ for
	 $Au+Au$ (solid line) and $Si+Si$ (dashed line; enhanced by a
         factor of 50) for 2.1 GeV/u; $\Theta_{CM} = 0^{o}$,
	 b = 0.

\item    \barp-multiplicity for $Si+Si$ at 2.1 GeV/u as a function of
	 the impact parameter b weighted with $2\pi b $; calculation:
	 solid line; analytic formula: dashed line (cf. text).

\item    Contributions to the production from different channels:
	 $NN$ (filed histograms), $N\Delta$ (open histograms) and
         $\Delta\Delta$ (hatched histograms) for $Si+Si$, $Ca+Ca$ and
	 $Au+Au$ at 2.0 GeV/u and 2.5 GeV/u.

\item    Relative contributions to the \barp-production from different
	 channels as a function of the beam energy for  $Ca+Ca$;
	 $NN$ (dotted line), $N\Delta$ (solid line) and $\Delta\Delta$
	 (dashed line).

\item    Differential \barp-production \prob for a central $Ni+Ni$
	 collision at 1.85 GeV/u ($\Theta_{CMS}=0^{o}$) as a
         function of the \apr momentum.\\
	 Top: fixed $U_{0}^{\bar{p}} = - 98$ MeV at $\rho=\rho_{0}$
	 and variing $U_{s}^{\bar{p}}$ at $\rho = \rho_{0}$: $-159$ MeV
	 (solid), $-100$ MeV (dashed), $-40$ MeV (dotted) and 0
	 (dashed-dotted).\\
	 Bottom: fixed $U_{s}^{\bar{p}} = - 159$ MeV at $\rho=\rho_{0}$
	 and variing $U_{0}^{\bar{p}}$ at $\rho = \rho_{0}$: $-98$ MeV
	 (solid), $-72$ MeV (dashed), $0$ (dotted) and $+72$ MeV
	 (dashed-dotted).

\item    Effects of the in-medium propagation on the \barp-sprectrum
         of $Si+Si$ at 2.1 GeV/u under $0^{o}$ in the CMS. \\
	 Top: $U_{s}^{\bar{p}} = 0$ and $U_{0}^{\bar{p}} = -72$ MeV
	 (dashed), 0 (solid) and $+72$ MeV (dotted) at $ \rho = \rho_{0}$.\\
         Bottom: $U^{\bar{p}}_{0}= 0$ and $U_{s}^{\bar{p}} = -100$ MeV
         (dashed), 0 (solid) and $+100$ MeV (dotted) at $\rho = \rho_{0}$.

\item    Effects of the in-medium propagation on the \barp-spectrum
	 for the reaction $p+Cu$ at 4.0 GeV, b= 0, $\Theta_{lab}=0^{o}$
	 for different optical potentials: $U_{sep}=0$ (dotted),
	 $U_{sep}=-90$ MeV (solid) and $U_{sep} = -560$ MeV (dashed).

\item	 Number of \apr-baryon collisions for a central $Si+Si$ reaction
         at 2.1 GeV/u as a function of the corresponding elementary
	 \barp-annihilation cross section.

\item    Antiproton spectra for the reaction $Si+Si$ at 2.1 GeV/u
         (b = 0, $\Theta_{CMS} = 0^{o}$) using different elementary
	 annihilation cross sections: upper line: no absorption; dotted
	 line: $\sigma_{abs}=60$ mb; dashed-dotted line: $\sigma_{abs} =
	 70$ mb; solid line: parametrization (\protect\ref{para});
	 dashed line: $\sigma_{abs}=80$ mb.

\item    Quotient R as a function of the impact parameter scaled with
	 $b_{max}$ for the systems $Ni+Ni$, $Ne+Ni$ and $Si+Si$ at
         1.85 GeV/u, 2.0 GeV/u and 2.1 GeV/u. Dots: RBUU-model;
         solid lines: fits according to eq. (\protect\ref{fit5}).

\item    Scalar and vector selfenergies $U_s(p)$ and $U_0(p)$ for nucleons
	 at different densities in units of $\rho_0 \approx 0.17 fm^{-3}$.

\item    Invariant cross section for antiproton production in the reactions
	 p+$^{12}C$ and
	 p+$^{63}$Cu at $\Theta=0^o$ as a function of the antiproton momentum
	 p in the lab. system. The experimental data are taken from
	 ref.\ \protect\cite{KEK} and correspond to bombarding energies
	 of 5.0 GeV , 4.0 GeV  and 3.5 GeV. The full lines represent
	 calculations for free antiprotons.
	 The dashed lines indicate the result for an antiproton selfenergy of
	 - 100 MeV at 4.0 GeV.

\item    Invariant cross section for antiproton production in the reaction
	 $^{28}Si+^{28}Si$ at 2.1 GeV/u and Ni + Ni at 1.85 GeV/u for
         $\Theta=0^o$ as a function of the momentum of the
	 emitted antiproton in the lab-system. The experimental data have been
	 taken from
         refs. \protect\cite{BEVALAC1} and \protect\cite{GSI}, respectively.
	 The upper lines indicate the calculated cross section for free
	 antiprotons without reabsorption whereas the lower solid line is
	 obtained when including $\bar{p}$ annihilation. The dashed line
	 represents the cross section adopting an attractive potential
	 of the antiproton of - 150 MeV.

\item	 Invariant cross section for antiproton production in the
	 reaction $Ni + Ni$ at 2.5 GeV/u (solid), 2.0 GeV/u (dashed),
	 1.85 GeV/u (dotted), 1.6 GeV/u (dotted-dashed) and
	 1.4 GeV/u (dotted-dotted-dashed). All calculations were done
         using an attractive \barp-potential of -100 MeV at $\rho =
         \rho_{0}$. The experimental data are taken from \protect\cite{GSI}.

\item    Comparison of our extracted values for the Schr\"odinger
         equivalent antiproton potential from p + A and A + A reactions
         with the prediction from the $\sigma - \omega$ model (dashed line)
         and the dispersive potential according to eqns. (\ref{dispses1}) -
         (\ref{cass5}) (solid line).

\end{list}
\end{appendix}

\newpage
\begin{table}
\begin{center}
\begin{tabular}
    {lllllll} \tableline
\multicolumn{4}{c}{2.0 GeV/u}
    &\multicolumn{3}{c}{2.5 GeV/u} \\
    \tableline
    &  $NN$
    &  $N\Delta$
    &  $\Delta\Delta$
     &  $NN$
    &  $N\Delta$
    &  $\Delta\Delta$
   \\ \hline
Si+Si & 5\, \% & 45 \,\% & 50 \,\% & 10 \,\%& 52 \,\% & 38\,\% \\
Ca+Ca & 6\, \% & 46 \,\% & 48 \,\% & 8 \,\% & 50 \,\% & 42\, \% \\
Au+Au & 4\, \% & 42 \,\% & 54 \,\% & 6 \,\%& 44 \,\% & 50\, \% \\
\end{tabular}
\end{center}
\caption{Relative contributions of the different reaction channels
integrated over $N_{1}+N_{2}$.}
\label{tab1}
\end{table}
\begin{table}
\begin{center}
\begin{tabular}
    {lrrr} \tableline
    & Ni + Ni
    & Ne + Ni
    & Si + Si
                        \\ \hline
A   & 1.0022    & 1.0025  & 1.0122  \\
B   & 0.9       & 0.9     & 0.9     \\
C   & 0.05      & 0.04    & 0.06    \\
\tableline
\end{tabular}
\end{center}
\caption{Parameters A, B and C for the fit (\protect{\ref{fit5}}) of $R$.}
\label{tab2}
\end{table}

\end{document}